\documentclass[pre,amsmath,amssymb,twocolumn,showpacs,showkeys]{revtex4}
\usepackage{graphicx}
\usepackage{dcolumn}
\usepackage{bm}
\usepackage{subfigure}
\usepackage{epsfig}

\begin{document}

\preprint{APS/123-QED}

\title{Efficiency of Rejection-Free Methods for 
Dynamic Monte Carlo Studies of Off-lattice Interacting Particles}

\author{Marta L. Guerra}
\author{M. A. Novotny}
\affiliation{Dept. of Physics and $HPC^2$ Center for Computational
    Sciences, P.O. Box 5167, Mississippi State University, 
    Mississippi State, MS, 39759, USA}
\author{Hiroshi Watanabe}
\affiliation{Department of Complex Systems Science, Graduate School of
  Information Science, Nagoya University, Furo-cho, Chikusa-ku, 
Nagoya 464-8601, Japan}
\author{Nobuyasu Ito}
\affiliation{Department of Applied Physics, University of Tokyo, Hongo 7-3-1, Bunkyo-ku, Tokyo 113-8656, Japan}

\date{\today}

\begin{abstract}
We calculate the efficiency of a rejection-free dynamic Monte Carlo method 
for $d$-dimensional off-lattice homogeneous 
particles interacting through a repulsive power-law potential $r^{-p}$. 
Theoretically we find the algorithmic efficiency in the limit of 
low temperatures and/or high densities
is asymptotically proportional to
$\rho^{\tfrac{p+2}{2}}T^{-\tfrac{d}{2}}$ with the particle density $\rho$
and the temperature $T$.
Dynamic Monte Carlo simulations are performed in 1-, 2- and 
3-dimensional systems
with different powers $p$, and the results agree with the 
theoretical predictions.
\end{abstract}

\maketitle

\section{Introduction}
Since their introduction in 1953 \cite{metropolis} 
classical Monte Carlo (MC) methods mave matured into 
a useful tool for studying many different
phenomena in different fields such as material science, high energy physics, 
and biology \cite{binder,miyashita,binder97}. 
For the study of the statics of a given model, 
the MC method can be viewed as a 
probabilistic method of performing multi-dimensional integrals \cite{press}
that could correspond to the partition (or grand partition) function 
of the model \cite{binder}.  Various methods to increase the efficiency 
or accuracy of 
the method exist, including importance sampling \cite{binder}, which 
allows the MC method to provide an estimate for the ratio of two 
integrals, for example to give an estimate of the average energy 
$\langle E\rangle$ of the system.  
Other advanced algorithms, such as Swendsen-Wang or cluster algorithms 
\cite{binder,landau}, can also be used to alleviate difficulties associated 
with critical slowing down near $T_c$ or being frozen into valleys 
for $T<T_c$.  All of these advanced methods are allowed because they 
provide estimates for the underlying integral(s) in an efficient fashion.  
For any of these MC methods, the system to be studied is in some 
configuration, which has a particular energy $E_i$, and an algorithm to 
obtain a new configuration $j$ from configuration $i$ is implemented.  For 
$M$ generated configurations, the estimate for the average energy is given by 
$\langle E\rangle = \frac{1}{M} \sum_{i=1}^M E_i$.  

If one is interested in the physical time evolution of a 
model system, the MC method can still be used.  Although in 
principle any MC algorithm for statics could be used to study 
the time development through phase space of the model system, only certain 
MC methods will correspond to the actual time evolution of the 
system being modeled.  In other words, many of the 
methods mentioned above, such as 
Swendsen-Wang or cluster algorithms \cite{binder,landau}, change the 
rate at which the system moves through phase space and consequently 
would usually not be associated with the actual time development 
of the physical system.  The older MC literature simply refers to the method 
as a MC method (see the references in Sec.~3.4 of Ref.~\cite{binder97}), even 
when the time development of the MC algorithm 
is assumed to correspond to that of the actual model system \cite{stoll}.  
More recently, use of MC methods to study the physical 
time dependence of a model system has been called either dynamic MC 
or kinetic MC.  Although these two terms are sometimes used 
interchangably, there is an emerging distinction between them.  
Kinetic MC has become the standard name 
for the case where physical time development is studied with known 
rate constants for the system to evolve from one state to 
another \cite{voter}.  
These rate constants may be approximated under 
certain assumptions (such as applicability of transition state theory to 
atomistic systems) using {\it ab initio\/} methods \cite{voter}.  
(Use of rate constants also allows the transition from 
discrete MC steps to continuous time.)
However, there are other instances where the physical time evolution of 
the system is desired while rate constants might be unavailable (for example 
perhaps transition state theory does not apply or an important complicated 
multi-particle motion might be difficult to conceptualize or calculate).
In such cases the physical time development may sometimes 
still be derived from the 
underlying physical system, for example by studying the underlying 
quantum mechanism for time development \cite{martin, park1, park, park2},
or devising a method equivalent to the time development
of the underlying equations \cite{okabe,cheng07}.

Such studies, which we concentrate 
on in this article, are called dynamic MC studies to distinguish 
them from static (equilibrium) MC or from kinetic MC studies.  
Thus we use the term dynamic MC in the same way as the 
recent book by Krauth \cite{krauth}.  

Frequently in dynamic simulations we need to work with long time 
scales at very low
temperatures or in a strong external field.  
In these cases the standard dynamic MC method becomes very inefficient due to
the high rejection rate which requires a large number of trial moves before a
change is made to the state of the system.
Most advanced algorithms, such as Swendsen-Wang or cluster algorithms 
\cite{binder,landau}, change the dynamic of the system thereby 
changing the time development of the system, 
which makes it impossible to study systems where the MC move 
is based on physical processes.  
The rejection-free MC (RFMC) method was proposed to overcome 
this problem with standard dynamic MC.
The RFMC was first applied to
discrete spin systems \cite{bortz, gillespie1, gillespie2} 
including the kinetic Ising Model
\cite{miyashita}.  
It was later generalized to classical spin systems with continuous degrees 
of freedom \cite{munoz}.  
The RFMC method allows us to efficiently 
simulate a system with a high rejection rate
without any changes of the original dynamics,
since it shares the original Markov chain with the standard 
dynamic MC method.  The RFMC method in this paper is very similar 
to the method labeled \lq faster-than-the-clock algorithm' in 
Ref.~\cite{krauth}.  

The RFMC requires, to proceed by one algorithmic step, the values of all the
probabilities of choosing a new state. Therefore, the 
computational cost of one step
is larger than that of the standard MC. For this reason 
it is necessary to have a method to
calculate its efficiency on a particular problem without 
implementing the method directly. Watanabe \textit{et al.}~developed a 
method \cite{watanabe} to calculate the
efficiency of the RFMC for spin systems and hard particle systems. 
In the present
paper, we evaluate the efficiency of the RFMC method for dynamic 
MC studies 
of $d$-dimensional particle systems with the 
particles interacting through a repulsive short-range power law potential.
Even though the bookkeeping involved in actually implementing the RFMC 
method may be substantial, leading to more computer time per algorithmic 
step than the standard dynamic MC method, at 
temperatures low enough or fields high enough where rejection rates 
are extremely high, the RFMC will be more efficient than the standard 
dynamic MC algorithm.  

The paper is organized as follows.
In Sec.~II, we present a review of the standard dynamic MC 
algorithm and the RFMC algorithm. In Sec.~III, we
provide analytical estimates for the efficiency of the RFMC method for 
repulsive power law potentials.
In Sec.~IV, we show the results of our simulations in 1, 2 and 3 dimensions.
Sec.~V is devoted to discussions and conclusions.

\section{Dynamic Monte Carlo}
\subsection{Standard Dynamic Monte Carlo}
The standard dynamic MC algorithm for particle systems involves the 
following six iterative steps, with one iteration being called a 
MC step (MCS).  We have used the term dynamic MC for 
this algorithm, rather than the term time-quantified MC used in 
\cite{okabe,cheng07}, since it has been used 
previously \cite{park1, park, park2,krauth}.  
It is important to remember that in dynamic MC the 
time in MCS is proportional to the physical time 
in seconds \cite{martin, park1, park, park2,okabe,cheng07}.  
The algorithm satisfies detailed balance.  It is very similar to the 
time-quantification of the dynamic MC for Brownian ratchets \cite{cheng07}.  
\begin{enumerate}
\item Choose one particle randomly from the $N$ particles, the chosen 
particle $i$ is located at position ${\vec r}_{{\rm old},i}$.
\item Choose a new position of the chosen particle randomly as
${\vec r}_{{\rm new},i}={\vec r}_{{\rm old},i}+\Delta {\vec r}$,
with $\Delta {\vec r}$ chosen uniformly over a
$d$-dimensional hyper-spherical volume
\begin{equation}\label{eq:2.Volume}
	V_{\rm choose} = \frac{\pi^{\frac{d}{2}}\> r_{\rm choose}^d}
	{\Gamma\left(\frac{d}{2}+1\right)}\>,	
\end{equation}
with a radius $r_{\rm choose}$ and the gamma function $\Gamma$.
The probability density for choosing the new
position of the chosen particle is $d^dx_i/V_{\rm choose}$.
\item Reject the new position if it is located outside the `cage' formed by
  line segments joining its nearest-neighbor (nn) particles.  
  (In the usual way, the `cage' is defined in this off-lattice 
  simulation on the basis of a Voronoi diagram 
  (or Delaunay triangulation), but 
  can be often equally well defined for our homogeneous high-density systems 
  as particles within a certain distance of the chosen atom \cite{watanabe}.)
\item Evaluate the energy difference 
  $\Delta E_i=E_{{\rm new},i}-E_{{\rm old},i}$
  between the new and the old positions of the chosen particle $i$.
\item Decide whether to accept the trial move by comparing a random
  number with the move probability which is a function of $\Delta E_i$.
  For example, we can use the Metropolis criteria to choose the 
  transition probability as
\begin{equation}\label{EqPi0}
P({\vec r}_{{\rm new},i}\mid {\vec r}_{{\rm old},i})=\left\{
\begin{array}{cc}
1 & \text{if~} \Delta E_i\leq 0,\\
\exp (-\beta\Delta E_i) & \text{otherwise}.  
\end{array}
\right.
\end{equation} 
\label{enu:Metropolis}

\item If the trial move is accepted, move the particle to its new position,  
  otherwise leave it in its old position.  
\end{enumerate}

\subsection{Dynamic Monte Carlo Without Rejections}
The rejection-free MC (RFMC) method was developed to overcome the decrease
in the efficiency of the standard dynamic MC method in cases 
where the rejection
rate is high, for example, in systems at a very low temperature. 
The efficiency of the standard dynamic MC algorithm is the
rate at which it changes the current state of the system, this is the fraction
of accepted moves to the total number of move attempts.
One algorithmic step of the RFMC involves the following procedures.
\begin{enumerate}
\item Compute the time to leave the current state
(the waiting time $t_{\text{wait}}$).
This is the number of trial states which would be rejected in the 
standard dynamic MC.
Hence in one algorithmic step the time is advanced by $t_{\text{wait}}$.
\item Advance the time of the system by $t_{\text{wait}}$.
\item Calculate probabilities $\lambda_i$ for each of the $N$ particles,
where $\lambda_i$ denotes the probability that the 
trial move of particle $i$ would be rejected in the standard dynamic MC 
given that it was the particle chosen for the trail move.  Explicitly 
\begin{equation}\label{EqLambdaI}
\lambda_i \> = \> 1 \>-\> 
\frac{1}{V_{\rm choose}} \int_{V_{\rm choose}}
P({\vec r}_{{\rm new},i}\mid {\vec r}_{{\rm old},i}) d^dx_{{\rm new},i}
.
\end{equation} 
\item Then choose a particle with the probability proportional to 
$1-\lambda_i$,
that is, the probability that the particle $i$ is chosen is given by
$\displaystyle \frac{1-\lambda_i}{\sum_{k=1}^N (1-\lambda_k)}$.
Therefore, the particle which is easy to move has a higher probability 
to be chosen.
\item Choose a new position of the chosen particle.
This is accomplished using the probability density 
\begin{equation}\label{EqLambdaChooseI}
\frac{P({\vec r}_{{\rm new},i}\mid {\vec r}_{{\rm old},i}) d^dx_{{\rm new},i}}
{V_{\rm choose}\left(1-\lambda_i\right)} 
.
\end{equation} 
\end{enumerate}
The above procedure does not contain any rejection step,
and therefore, each RFMC algorithmic 
step always involves a change of state of the system.

The efficiency of the RFMC method is inversely proportional to the rejection
rate of the standard dynamic MC, but they share the same dynamics. 
Therefore the efficiency of the RFMC is related to the inefficiency of 
the standard dynamic MC. For
this reason the efficiency of the RFMC will be proportional to
$t_{\text{wait}}$ which is given by \cite{watanabe,novotny95}:
\begin{equation}\label{Eqtwait}
    t_{\text{wait}}=\biggl\lfloor\frac{\ln \tilde r}
    {\ln \Lambda}\biggr\rfloor+1.
\end{equation}
Here $\tilde r$ is a random number uniformly distributed on $(0,1]$,
  $\lfloor\cdot\rfloor$ is the integer part, and 
  $\Lambda=\frac{1}{N}\sum_{i=1}^N \lambda_i$ 
is the probability
to stay in the current state (that the move to the trial state will be 
rejected) after one standard dynamic MC trial move.  
The units of time are in MCS, but can be 
quantified with physical time 
\cite{martin, park1, park, park2,okabe,cheng07}.  
We have used the Metropolis method \cite{metropolis}
as shown in Eq.~(\ref{EqPi0}) in this 
example and in our simulations in the next section.  
However, the RFMC algorithm would also work for different functional
probabilities such as the Glauber or heat-bath dynamic 
\cite{binder,miyashita,martin} or a phonon dynamic \cite{park1, park2}.

\section{Efficiency of RFMC for Power-law Potentials}
Consider $d$-dimensional particles with a repulsive power-law potential
and $N_{\rm nn}$ nearest-neighbors (nn) [the particles that form 
its `cage']. The potential between any two nn
particles is
\begin{equation}\label{EqPLpotential}
V(r)=\left\{
\begin{array}{cc}
\Bigl(\displaystyle\frac{\sigma}{r} 
\Bigr)^p-\Bigl(\displaystyle\frac{\sigma}{r_0} 
   \Bigr)^p & r\leq r_0\\
0 & r \geq r_0
\end{array}
\right.
\end{equation}
where $r$ is the distance between particles, $p$ is the power,
$r_0$ is the cut-off distance and $\sigma$ is the length that gives 
the strength of the interaction, respectively.
This potential function represents the hard repulsive core 
of any potential, such as the Lennard-Jones potential which has $p=12$, 
and this repulsive part becomes dominant at high densities.

We have chosen the origin of the coordinate system to
be at the position of the chosen atom $i$.
The energy difference of the atom $i$ is given 
by $\Delta E_i = U_i(\vec x)-U_i(\vec 0)$ with $\vec x$ the trial position.
The efficiency of the rejection-free
algorithm at low temperatures and/or high densities is given 
for the Metropolis dynamic by
\begin{widetext}
\begin{equation}\label{eq:2.E1}
	\bigl\langle \exp\left[-\beta \Delta E\right]\bigr\rangle =
	\frac{\Gamma\left(\frac{d}{2}+1\right)} 
            {\pi^{\frac{d}{2}}\> r_{\rm choose}^d}
	\int_{-\infty}^\infty\cdots\int_{-\infty}^\infty d^dx \>
	\Theta_{\rm cage} \> \exp\left\{-\beta\left[
	U_i(\vec x)-U_i(\vec 0)
	\right]\right\},
\end{equation}
\end{widetext}
where the interaction energy for particle $i$ is given by the power-law
dependence, $p$, of the repulsive part of the interatomic potential
\begin{equation}\label{eq:2.EU}
	U_i\left({\vec x}\right)= \sum\limits_{N_{\rm nn}} V(r)= 
	\sum_{k=1}^{N_{\rm nn}} \frac{\sigma^p}
	{|\vec x- \vec x_k|^p},
	\end{equation}
with $\vec x_k$ the position of the $k^{th}$ nn atom of
the chosen atom $i$. Here $\Theta_{\rm cage}$ restricts the integrand to be
non zero only with the cage formed by the nn particles. The angular brackets
denote an average over all allowed states of the system weighted with the
Boltzman weight at each configuration.  
Since we are interested in the system at high densities or at low 
temperatures, we can utilize the Laplace saddle-point 
integration approximation \cite{saddle}
\begin{equation}\label{eq:saddle}
  Z_p=\int\cdots\int d^dxP(\vec x)\approx P(\vec x_0)\sqrt{\frac{(2\pi)^d}
    {{\rm det}{\bf A}}}
\end{equation} 
where the integrand $P(\vec x)$ is strongly peaked around $\vec x=\vec x_0$
and
\begin{equation}\label{eq:matrixA}
  A_{ij}=-\frac{\partial^2}{\partial x_i\partial x_j}\biggl. \ln\left[P(\vec
    x)\right]\biggr|_{\vec x=\vec x_0}.
\end{equation}
We assumed the chosen particle is at or near its local 
energy minimum, \textit{i.e.}, $P\left({\vec x_0}\right)\approx 1$.
Therefore
\begin{equation}\label{eq:2.E2}
	\bigl\langle \exp\left[-\beta \Delta E\right]\bigr\rangle \approx 
	\frac{\Gamma\left(\frac{d}{2}+1\right)\>T^{\frac{d}{2}}}
	{\pi^{\frac{d}{2}}\> r_{\rm choose}^d}
	\sqrt{\frac{\left(2\pi\right)^d}{\left|{\rm det}\>
          {\tilde{\bf A}}\right|}},
\end{equation}
since after making the derivation in Eq.~(\ref{eq:matrixA}) we 
get a factor of $\beta$
and the only values that are not 0 are when $i=j$ we have defined 
$A_{ij}=\beta{\tilde A}_{ij}$, and the
determinant of ${\bf A}$ is thus proportional to $\beta^d$. This immediately
gives that the temperature dependence of the average waiting time is
proportional to $\beta^{\frac{d}{2}}=T^{-\frac{d}{2}}$.  

Furthermore, because of the power-law approximation of 
Eq.~(\ref{eq:2.EU}), and
that two partial derivatives must be taken for the saddle-point 
approximation $A_{ij}\sim r_{\rm nn}^{-p-2}$ where $r_{\rm nn}$ is the nn
distance if all nn atoms are equidistant from 
the chosen atom.  The particle density $\rho$ 
is proportional to $r_{\rm nn}^{-d}$,
and therefore, $\left|{\bf A}\right|^{-\frac{d}{2}} \sim 
\left[r_{\rm nn}^{-p-2}\right]^{-\frac{d}{2}}\sim \rho^{\frac{-p-2}{2}}$.
Equation~(\ref{eq:2.E1}), therefore becomes 
\begin{equation}\label{eq:2.E4}
	\bigl\langle \exp\left[-\beta \Delta E\right]\bigr\rangle \sim 
	\frac{T^{\frac{d}{2}}}
	{r_{\rm choose}^d\> \rho^{\frac{p+2}{2}}},
\end{equation}
and the average time between acceptances in the dynamic 
MC procedure is
\begin{equation}\label{eq:2.E5}
	\left\langle t_{\rm wait} \right\rangle \approx
	\frac{1}{\left\langle \exp\left[-\beta \Delta E\right]\right\rangle}
	\sim \frac{r_{\rm choose}^d\> \rho^{\frac{p+2}{2}}}{T^{\frac{d}{2}}}.
\end{equation}
Equation~(\ref{eq:2.E5}) is the main result of this paper. The result is very
general, both for various dimensional particles and for 
various power laws, as well as being general for the explicit dynamic that 
is used in the MC procedure. 

Throughout the present paper, we assume all the atoms have 
identical potentials.
In the following, we calculate the explicit expression of Eq.~(\ref{eq:2.E5})
for several conditions.
For the 1-dimensional system, 
the average waiting time is given by
\begin{equation}\label{eq:tw.1d}
	\left\langle t_{\rm wait} \right\rangle_{d=1} =\frac
	{2 r_{\rm choose}}{r_{\rm nn}^{(p+2)/2}}
          \sqrt{\frac{\sigma^p p (p+1)}{T\>\pi}}\sim
	\frac{\rho^{\frac{p+2}{2}}}{\sqrt{T}},
\end{equation}
since $V_{\rm choose}=2 r_{\rm choose}$, $N_{\rm nn}=2$, and 
\begin{equation}\label{eq:tw.1d.det}
	\left|{\tilde{\bf A}}\right|=\left|\frac{\partial^2 
	\left[U_i({\vec x})-U_i({\vec 0})\right]}{\partial x^2}\right|_{x=0} =
	\frac{2\sigma^p p\left(p+1\right)}{r_{\rm nn}^{p+2}}.
\end{equation}
For the 2-dimensional system which has a hexagonal 
lattice as the ground state, the average waiting time is given by
\begin{equation}\label{eq:tw.2d}
	\left\langle t_{\rm wait} \right\rangle_{d=2}=\frac
	{3\sigma^p p^2 r_{\rm choose}^2}{2 T r_{\rm nn}^{p+2}}\sim
	\frac{\rho^{\frac{p+2}{2}}}{T},
\end{equation}
since $V_{\rm choose}=\pi r_{\rm choose}^2$, $N_{\rm nn}=6$, and 
\begin{equation}\label{eq:tw.2d.det}
	\tilde A_{ij}=\left.\frac{\partial^2 
	\left[U_i({\vec x})-U_i({\vec 0})\right]}
	{\partial x_i\partial x_j}\right|_{{\vec x}=0} =-\delta_{ij}
	\frac{3\sigma^p p^2}{r_{\rm nn}^{p+2}},
\end{equation}
with the Kronecker delta $\delta_{ij}$.
For the 3-dimensional system which has the face-centered-cubic 
(FCC) lattice as the ground state,
the average waiting time is given by
\begin{equation}\label{eq:tw.3d}
	\left\langle t_{\rm wait} \right\rangle_{d=3} =
	\frac{\left[\sigma^p p (p-1)\right]^{\frac{3}{2}} \> r_{\rm choose}^3}
	{3 \sqrt{\pi} 2^{\frac{10+3p}{4}} 
	T^{\frac{3}{2}} r_{\rm nn}^{\frac{3(p+2)}{2}}}
	\sim\frac{\rho^{\frac{p+2}{2}}}{T^{\frac{3}{2}}},
\end{equation}
since  $V_{\rm choose}=(4/3)\pi r_{\rm choose}^3$, $N_{\rm nn}=12$, and 
\begin{equation}\label{eq:tw.3d.det}
	\tilde A_{ij}=\left.\frac{\partial^2 
	\left[U_i({\vec x})-U_i({\vec 0})\right]}
	{\partial x_i\partial x_j}\right|_{{\vec x}=0} =-\delta_{ij}
	\frac{\sigma^p p (p-1) 2^{2+\frac{p}{2}}}{r_{\rm nn}^{p+2}}.
\end{equation}
Note that, the density and the temperature dependence 
in a simple-cubic lattice is equivalent to that of 
Eq.~(\ref{eq:tw.3d.det}),
while the coefficient is different since 
$N_{\rm nn}=6$ and 
$\tilde A_{ij}= -\delta_{ij} \left(\sigma^p p (p-1)\right)/r_{\rm nn}^{p+2}$.

\section{Simulations in $d$ = 1, 2, and 3}
Following the methodology described above, we performed simulations for
1-, 2-, and 3-dimensional systems.
The goal is to locate where the asymptotic results of Eq.~(\ref{eq:2.E5}) 
hold for our RFMC method.  
The density of the system $\rho$ is defined to be $\rho\equiv
N(2a/L)^d$, with the number of particles $N$, the radius of the particles $a$,
the linear system size $L$, and the dimensionality of the 
system $d$, respectively.
Throughout all the simulations $r_{\text{choose}}$ is set to 0.05 
and the cut-off radius
$r_0$ is set to $r_0=1.1r_{nn}(\rho)$, with $r_{nn}(\rho)$ the
nearest-neighbor distance for the given density. The simulations were
performed for four values of $p$: 2, 4, 6, and 12. 
For all values of $d$, the density is 
fixed at $\rho=2.0$
in order to study the temperature dependence, the temperature 
is fixed at $T=0.001$
to study the density dependence.  All simulations were performed starting 
from the ground state, with periodic boundary conditions, with the 
simulated volume such that an 
integer number of unit cells of the ground state fit into the volume.  

\begin{figure}[hbt]
\begin{center}
\subfigure[]{\label{twaitVsTPL1D}
\includegraphics[width=6cm]{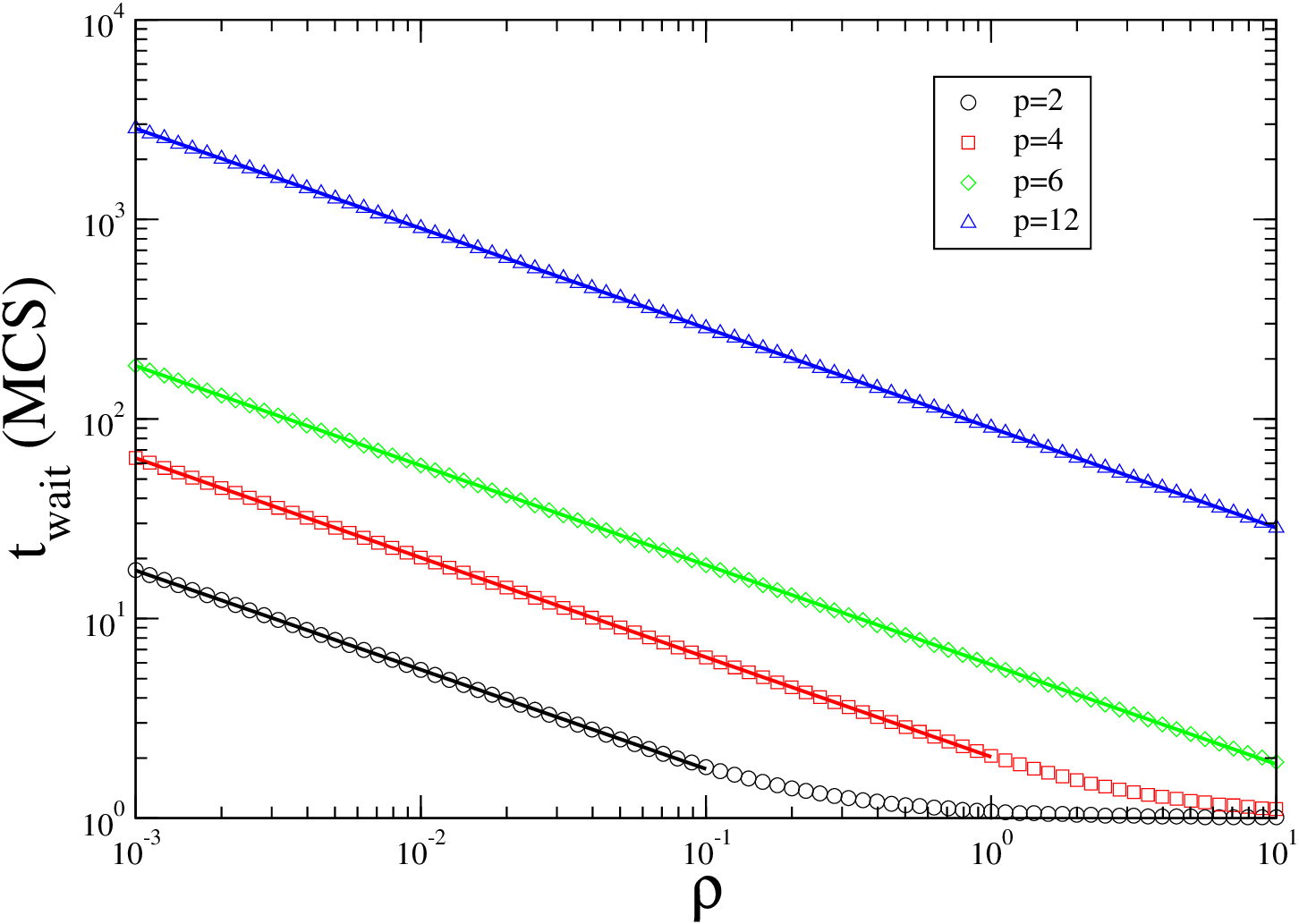}}\\
\subfigure[]{\label{twaitVsdenPL1D}
\includegraphics[width=6cm]{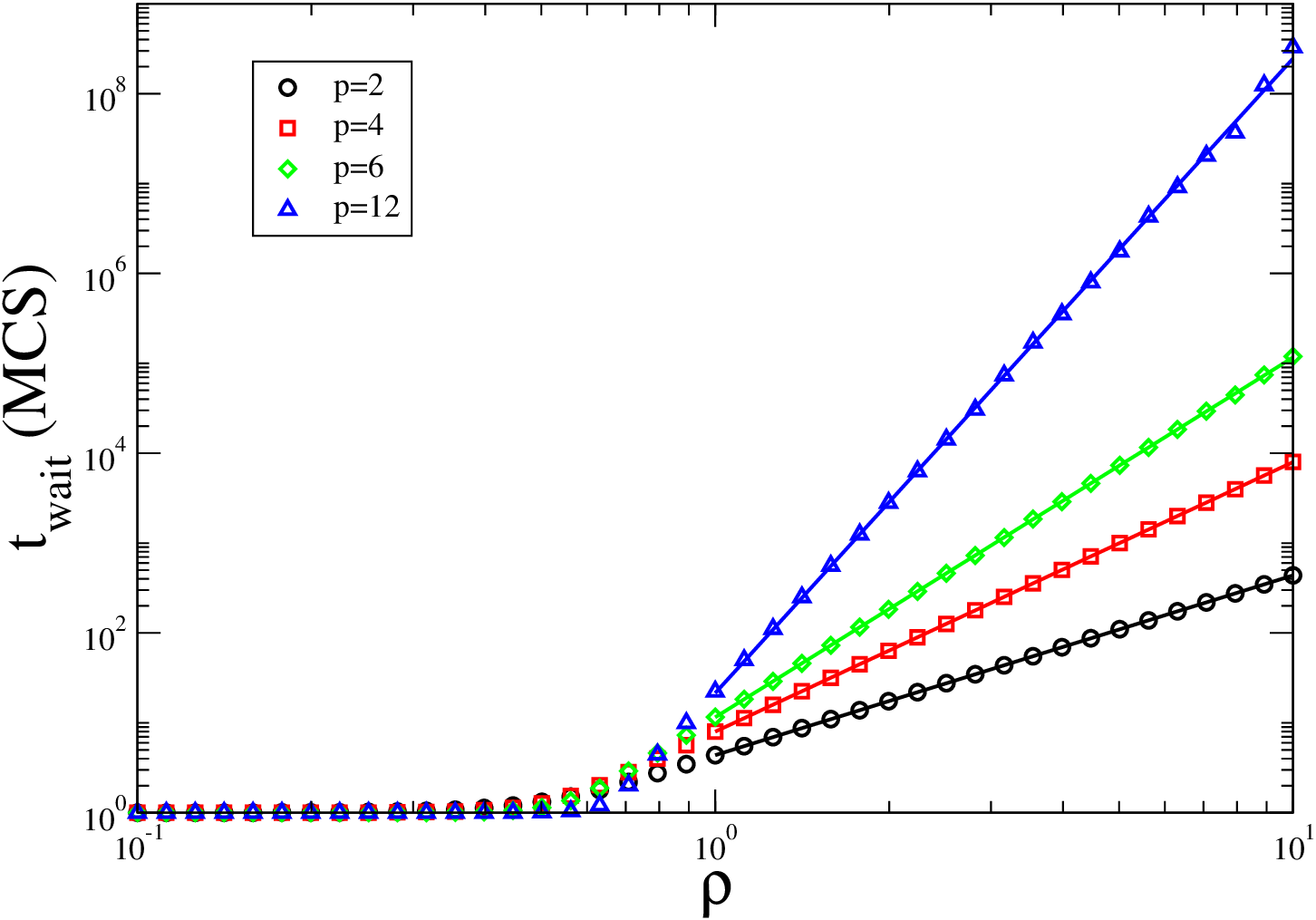}}
\caption{(Color online) (a) Temperature dependence of the average 
$t_{\text{wait}}$ in $d=1$
for $N=200$ particles, the solid lines are the power law fits. (b) Density
dependence of $t_{\text{wait}}$ in $d=1$, the solid lines are power law
fits with $\rho\ge 0.6$.}\label{twait1D}
\end{center}
\end{figure}

We study the 1-dimensional system with $N=200$ particles which are 
located on the line
of length $L$ with periodic boundary conditions. The temperature dependence of
the average for 
$t_{\text{wait}}$ is shown in Fig.~\ref{twait1D}~(a), in this case 
with statistics from $10^6$
MCS per particle (MCSp). 
The power law fit gives $\langle t_{\text{wait}}\rangle\sim T^{-0.49(1)}$ with
the correlation coefficient $r=1.00(1)$ for $p=2, 4$ and 6, while for $p=12$ 
we obtain $\langle t_{\text{wait}}\rangle\sim T^{-0.50(1)}$
with $r=1.00(1)$.  
We have fit the region with 
$T\le 10^{-1}$, $T\le 10^{0}$, $T\le 10^{1}$, and $T\le 10^{1}$ 
for respectively $p=2$, $4$, $6$, and $12$.  
All of these results are in agreement with our 
prediction, \textit{i.e.},
$\langle t_{\text{wait}}\rangle\sim T^{-0.5}$. The density dependence 
of $t_{\text{wait}}$ is
shown in Fig.~\ref{twait1D}~(b) with the number of trials 
in this case 
$10^6$ MCSp, or $10^7$ MCSp for high density and $p=12$. 
The power law fit in
this case, all for $\rho\ge 0.6$, for 
$p=2$ gives $\langle t_{\text{wait}}\rangle \sim
\rho^{1.99(1)}$ with $r=-1.00(1)$, for 
$p=4$ gives $\langle t_{\text{wait}}\rangle \sim
\rho^{2.99(1)}$ with $r=-1.00(1)$, for $p=6$ gives 
$\langle t_{\text{wait}}\rangle \sim \rho^{4.00(1)}$ with $r=-1.00(1)$, 
and for $p=12$ gives $\langle
t_{\text{wait}}\rangle \sim \rho^{7.00(1)}$ with $r=-1.00(1)$. Again our 
results agree with our asymptotic predictions, \textit{i.e.}, 
$\langle t_{\text{wait}}\rangle \sim \rho^{(p+2)/2}$. 
\begin{figure}[hbt]
\begin{center}
\subfigure[]{\label{twaitVsTPL2D}
\includegraphics[width=6cm]{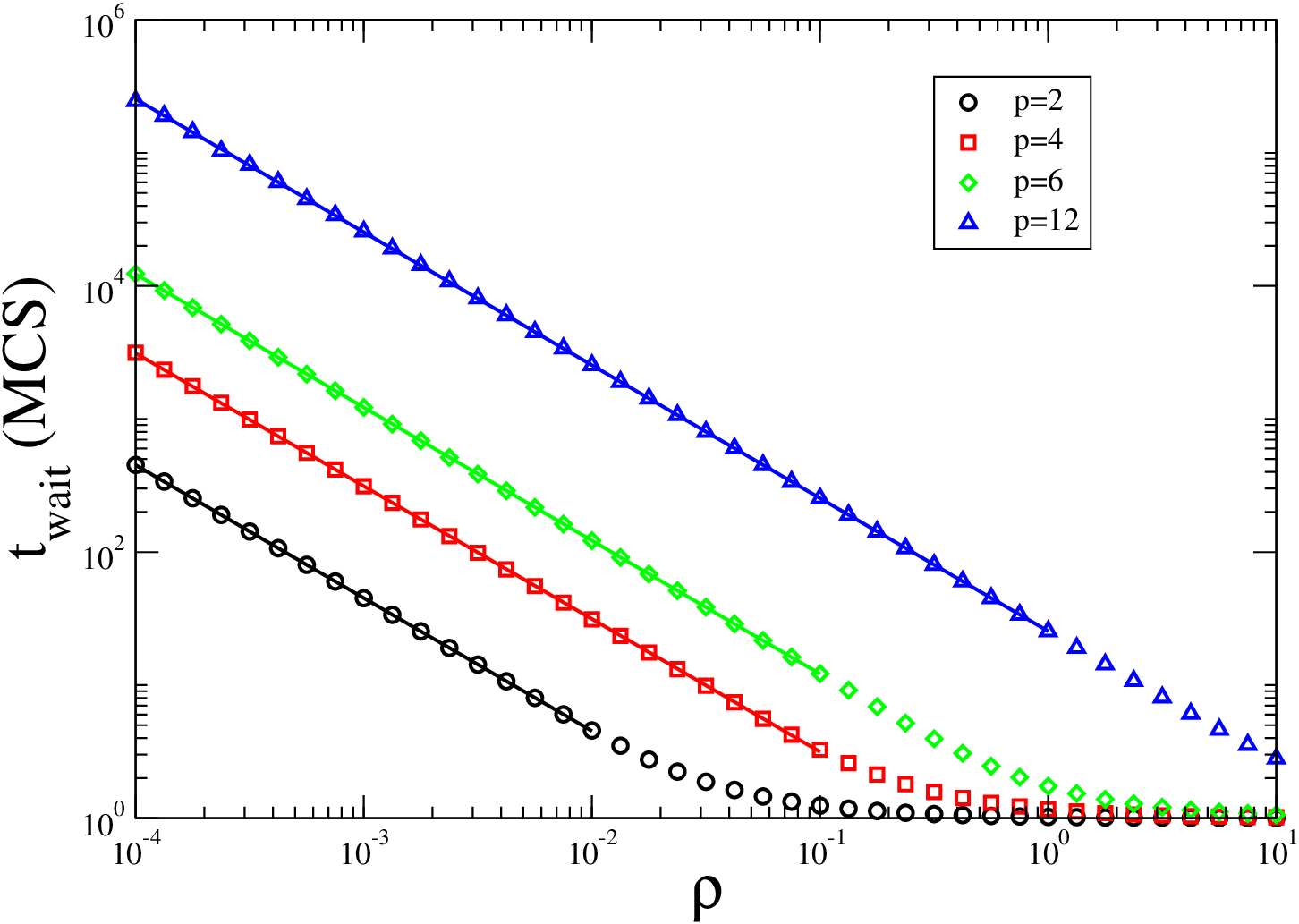}}\\
\subfigure[]{\label{twaitVsdenPL2D}
\includegraphics[width=6cm]{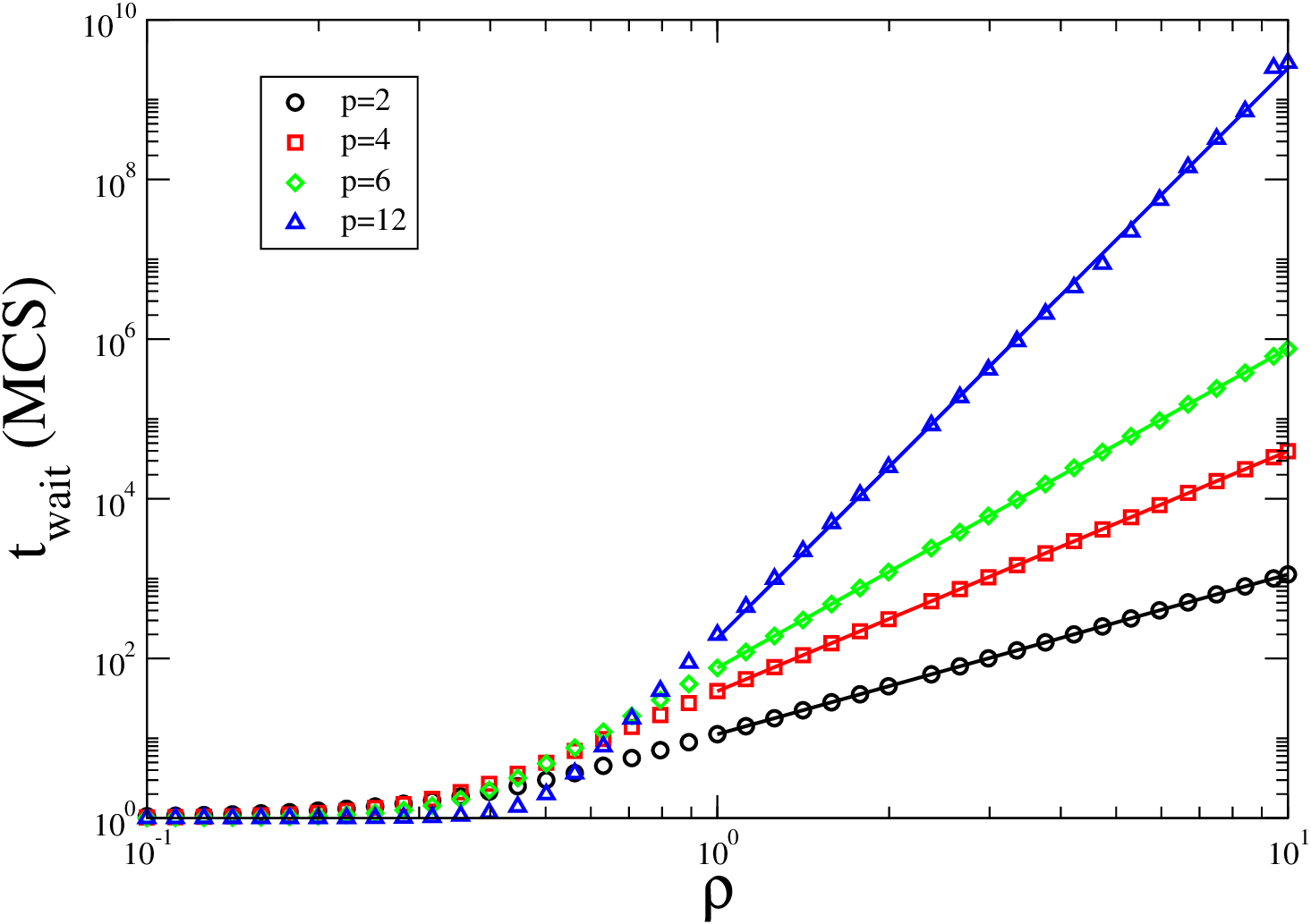}}
\caption{(Color online) (a) Temperature dependence and (b) density dependence
of the average $t_{\text{wait}}$ in $d=2$ for $N=80$ particles. 
The solid lines are the power law fits.
}\label{twait2D}
\end{center}
\end{figure}

We study the two-dimensional system with $N=80$ particles distributed on a 
triangular lattice of length
$L$ and width $\sqrt3 L/2$ 
with periodic boundary conditions for both axes.
Fig.~\ref{twait2D}~(a) shows the
temperature dependence of the average 
$t_{\text{wait}}$, in this case with the 
number of samples
$10^7$ MCSp. The power law fits, all for $\rho\ge 1$, 
give $\langle t_{\text{wait}}\rangle\sim
T^{-0.99(1)}$ with $r=1.0(1)$, for $p=2,4,6$ and 12. The density dependence 
is shown in Fig.~\ref{twait2D}~(b),
$10^6$ MCSp are taken for $p=2, 4, 6$ and $10^9$ for $p=12$ at high densities.
The power law fit for $p=2$ gives 
$\langle t_{\text{wait}}\rangle\sim\rho^{1.99(1)}$ with $r=-1.00(1)$, 
for $p=4$ gives 
$\langle t_{\text{wait}}\rangle\sim\rho^{3.00(1)}$ with $r=-1.00(1)$, 
for $p=6$ gives 
$\langle t_{\text{wait}}\rangle\sim\rho^{4.00(1)}$ with $r=-1.00(1)$, 
and for $p=12$ gives 
$\langle t_{\text{wait}}\rangle\sim\rho^{7.09(1)}$ with $r=-1.00(1)$.
These results show excellent agreement with the asymptotic prediction 
$\langle t_{\text{wait}}\rangle\sim\>T\>\rho^{\frac{p+2}{2}}$
in Eq.~(\ref{eq:tw.2d}).

\begin{figure}[hbt]
\begin{center}
\subfigure[]{\label{twaitVsTPL3D}
\includegraphics[width=6cm]{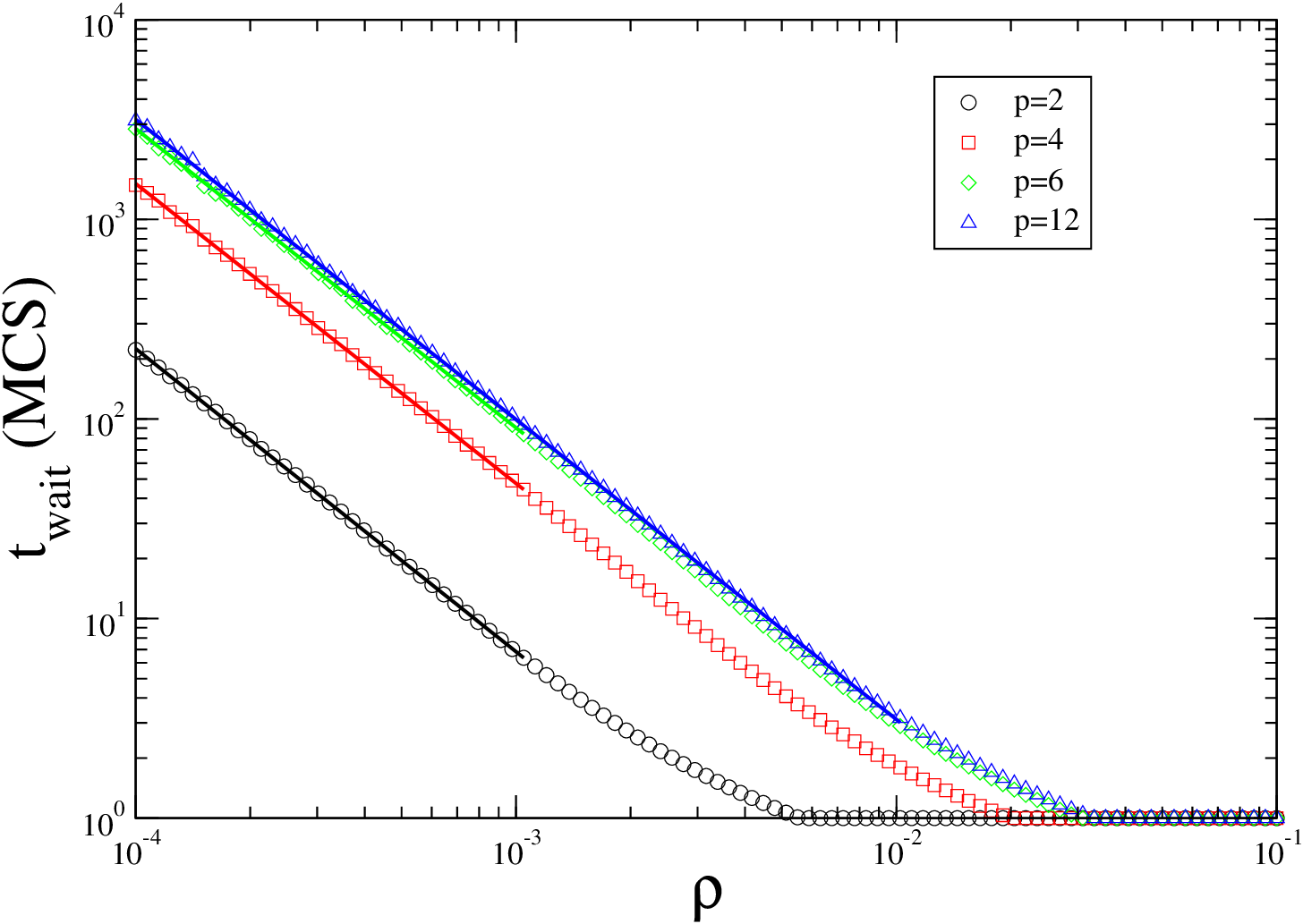}}\\
\subfigure[]{\label{twaitVsdenPL3D}
\includegraphics[width=6cm]{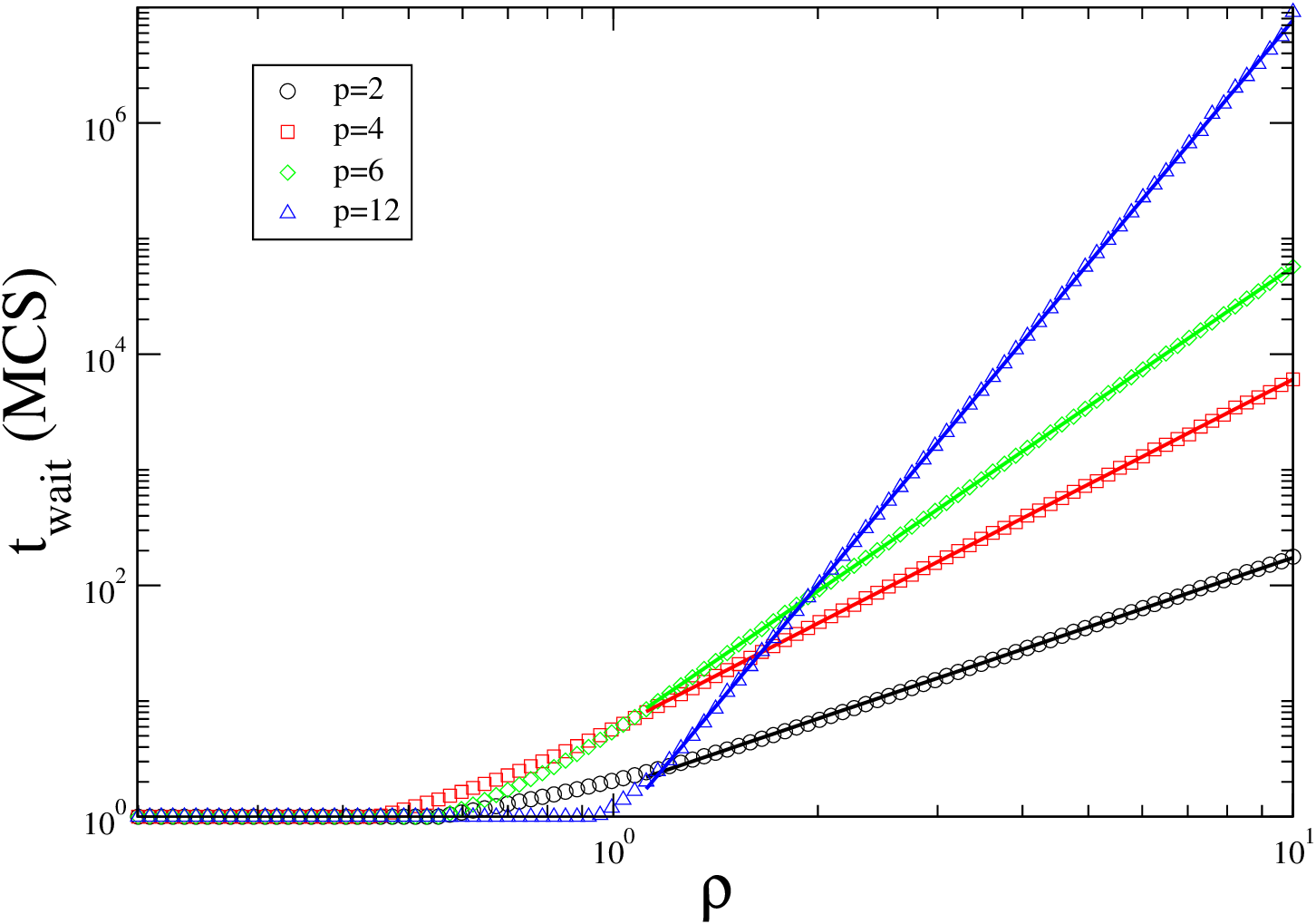}}
\caption{(Color online)
(a) Temperature dependence and (b) density dependence
of the average $t_{\text{wait}}$ for a 3-dimensional FCC system with 
$N=500$ particles.
The solid lines are the power law fits.
}\label{twait3D}
\end{center}
\end{figure}

We also study the 3-dimensional system with $N=500$ particles located on an 
FCC lattice with periodic boundary conditions for all directions.
The temperature dependence and density 
dependence of the average $t_{\text{wait}}$ are shown
in Fig.~\ref{twait3D}~(a) and Fig.~\ref{twait3D}~(b), respectively.
The power law fits of the temperature dependence gives $\langle
t_{\text{wait}}\rangle\sim T^{-1.49(1)}$ with $r=1.00(1)$, for 
$p=2, 4, 6$ and 12. The power
law fit, all for $\rho\ge 1.1$, for the density dependence for $p=2$ gives 
$\langle t_{\text{wait}}\rangle\sim\rho^{1.98(1)}$ with $r=-1.00(1)$, 
for $p=4$ gives 
$\langle t_{\text{wait}}\rangle\sim\rho^{3.02(1)}$ with $r=-1.00(1)$, 
for $p=6$ gives 
$\langle t_{\text{wait}}\rangle\sim\rho^{4.01(1)}$ with $r=-1.00(1)$, 
and for $p=12$ gives 
$\langle t_{\text{wait}}\rangle\sim\rho^{6.96(1)}$ with $r=-1.00(1)$.
The results agree with the asymptotic predictions 
$\langle t_{\text{wait}}\rangle\sim T^{-3/2}\rho^{(p+2)/2}$.
\section{Summary and Discussion}

We studied the efficiency of the Rejection Free Monte Carlo (RFMC) 
method for systems having particles 
interacting through repulsive power-law potentials.
The density and temperature dependence of the average waiting time has been 
predicted to be,
\begin{equation}
\langle t_{\text{wait}}\rangle \sim 
\frac{\rho^{\tfrac{p+2}{2}}}{T^{\tfrac{d}{2}}},
\end{equation}
with the dimensionality of the system $d$, density $\rho$, and the 
temperature $T$, respectively.  
These theoretical results are valid asymptotically for large $\rho$ and/or 
low $T$.  
Monte Carlo simulations were performed and the results showed good agreement 
with the asymptotic prediction. 
This study shows how efficient the RFMC method is 
in low temperature or in high density regimes.
Assume the wall-clock time per algorithmic step for the standard dynamic 
Monte Carlo algorithm is $t_0$, and the 
average wall-clock time per RFMC algorithmic step is $t_1$
(both of which are expected to be almost independent of $T$ and $\rho$).  
Because of the extra bookkeeping involved in programming 
the RFMC method, $t_1>t_0$.  Nevertheless, the RFMC method will be 
more efficient (use less wall-clock time) whenever 
$t_1 < t_0 \langle t_{\rm wait}\rangle$.  This inequality will 
always be satisfied for low enough $T$ or high enough density.

The RFMC method does not change the dynamic of the MC move associated with the
underlying physical dynamics, and therefore 
makes possible the study of systems with a fixed physical dynamic.
It is very important keeping the dynamics unchanged, since 
the change in the dynamics of the MC move can cause a strong influence
in certain dynamic 
physical properties~\cite{park, buendia, gillespie1, gillespie2}.

Although we studied a repulsive core 
power-law potential, we expect the equivalent
behavior of the waiting time for more realistic potentials, as long as 
we work with the system at low
temperatures and/or high densities. A further avenue of study, 
therefore, could be
to calculate the efficiency of the RFMC method for more 
general potentials such as
Lennard-Jones, or those derived from density functional theory. 
A related study could also be of the efficiency of the
RFMC in the case of two or more types of particles.  
The actual implementation and utilization of the RFMC method in particle 
simulations can be attempted now that the ultimate behavior of the 
algorithmic efficiency has been determined.  

\begin{acknowledgments}
The authors thank P. A. Rikvold for helpful discussions. The computation was
partially carried out 
in the ISSP of the University of Tokyo and in the 
$HPC^2$ Center for Computational Sciences at
Mississippi State University. 
This work was partially supported by 
the COE program on ``Frontiers of computational science'' of 
Nagoya University, 
U.S. NSF grants DMR-0426488 and DMR-0444051, 
the Sustainable Energy Research Center at Mississippi State University
which is supported by the Department of Energy under 
Award Number DE-FG3606GO86025, 
by Grants-in-Aid for Scientific Research (Contracts No.\ 19740235 
and 19540400), and by KAUST GRP(KUK-I1-005-04).  
\end{acknowledgments}
\bibliography{RFMCoffV3}
\end{document}